\documentclass[aps,pra,twocolumn,superscriptaddress,amsmath,amssymb,floatfix]{revtex4-2}

\usepackage{graphicx}
\usepackage{bm}
\usepackage{mathtools}
\usepackage{xcolor}
\usepackage{hyperref}
\usepackage{physics}
\usepackage{braket}
\usepackage{booktabs}

\newcommand{\ii}{\mathrm{i}}

\begin{document}

\title{Quantum Pump Depletion and Multicomponent Schr\"odinger-Cat-Like States
in Doubly Pumped Intraresonance Kerr Microresonators}

\author{Ranjit Singh}
\email{ranjit.singh@mail.ru}
\affiliation{Independent Researcher, Domodedovo, Moscow Region, Russia}

\author{Alexander E. Teretenkov}
\email{taemsu@mail.ru}
\affiliation{Department of Mathematical Methods for Quantum Technologies, Steklov Mathematical Institute of Russian Academy of Sciences, Moscow, Russia}

\date{\today}

\begin{abstract}
We investigate quantum pump depletion and non-Gaussian state generation in doubly pumped Kerr microresonators operating in the intraresonance regime. The pump modes are treated quantum mechanically rather than as undepleted classical amplitudes, allowing pump depletion, back-action, entanglement generation, quadrature fluctuations, and Wigner-function negativity to emerge from the same multimode dynamics. Starting from the Kerr four-wave-mixing selection rule, we distinguish an effective resonant photon-conversion model from the full Kerr Hamiltonian containing self-phase modulation (SPM), cross-phase modulation (XPM), and four-wave mixing (FWM). The reduced model isolates the photon-conversion network responsible for the discrete $\mathbb{Z}_{n+1}$ phase structure, whereas the full model retains operator-valued nonlinear Kerr phases. For the \(n=2\) intraresonance branch, the four-mode reduced initial-value problem with fixed coherent pump phases has a residual \(\mathbb{Z}_3\) symmetry and generates cat-like Wigner structures near the interaction length at which the generated-mode population \(\langle n_1\rangle\) is maximal and the pump population \(\langle n_0\rangle\) is strongly depleted. The resulting states are not the canonical even or odd coherent states of Dodonov, Malkin, and Man'ko, but multicomponent Schr\"odinger-cat-like states characterized by Wigner negativity, non-Poissonian statistics, pump-mode quadrature squeezing, and large single-mode Schmidt numbers. Comparison of the reduced and full Kerr dynamics shows that uncompensated SPM/XPM-induced phase shearing suppresses the interference fringes and Wigner negativity responsible for the clearest cat-like signatures. These results identify quantum-depleted intraresonance Kerr dynamics as a route to symmetry-organized non-Gaussian states in Kerr resonators.
\end{abstract}

\maketitle

\section{Introduction}

Optical Kerr microresonators provide a compact platform for nonlinear optics, frequency-comb generation, and quantum-state engineering. In conventional Kerr-comb generation, different spectral lines are associated with distinct cavity resonances separated by the free spectral range. A recently demonstrated alternative is the intraresonance regime, where two closely spaced pump tones drive one and the same optical resonance and generate additional frequency components inside the linewidth of that resonance~\cite{Intraresonance2026}. In this regime the generated components are separated by subharmonics of the dual-pump separation rather than by the resonator free spectral range.

The classical intraresonance comb picture predicts that when \(n\) intermediate lines are generated between two pumps, the field can possess \(n+1\) phase-stable branches. This is naturally connected with phase locking and synchronization phenomena in nonlinear dynamical systems~\cite{Alexandrov2026}. In the present quantum problem, the same phase-locking structure appears as a discrete \(\mathbb{Z}_{n+1}\) symmetry of the resonant photon-conversion network. The symmetry fixes the number of phase-related branches, while the quantum dynamics determines whether coherent interference between these branches survives in the reduced states.

Schr\"odinger-cat and cat-like states are useful resources in continuous-variable quantum information because coherent superpositions of well-separated phase-space components can encode bosonic qubits and protect information against specific noise channels~\cite{Cochrane1999,Mirrahimi2014}. Their interference fringes are also sensitive to weak displacements and phase-space perturbations, making squeezed cat states relevant for quantum sensing~\cite{SinghTeretenkovPhysicsOpen2024}. More generally, finite superpositions of coherent states provide a standard
phase-space model of nonclassical cat-like states: their Wigner functions
display interference fringes, photon-number oscillations, and strong
sensitivity to dissipation~\cite{Buzek1992}. Related depleted-pump studies of entangled Schr\"odinger-cat-like states in parametric down-conversion have also shown that Wigner-function interference, squeezing, super-Poissonian statistics, and entanglement can coexist in fully quantum nonlinear optical dynamics~\cite{SinghTeretenkovPLA2026}. In this context, the generation of non-Gaussian cat-like states inside integrated Kerr photonic platforms is of direct interest for continuous-variable computation, optical cat qubits, and sensing protocols.

For comparison, Appendix~\ref{app:undepleted} briefly examines the undepleted-pump limit, where the pump modes are replaced by fixed classical amplitudes. In that case, non-Gaussian states can still arise from the cubic photon-conversion processes acting on the quantum generated modes, but pump depletion, quantum back-action, and pump-generated-mode entanglement are absent by construction.

The central physical mechanism studied here is quantum pump depletion. In contrast to undepleted-pump descriptions, the pump modes are retained as quantum harmonic oscillators. Photon conversion therefore depletes the pump sector, generates back-action, and entangles the generated modes with the pumps. Thus, a fully quantum treatment of the pump modes is essential for a consistent description of non-Gaussian reduced-state generation with pump depletion, quantum back-action, and pump-generated-mode entanglement.

The states discussed in this work should be called Schr\"odinger-cat-like states rather than canonical even or odd coherent states. The canonical even and odd coherent states introduced by Dodonov, Malkin, and Man'ko are characterized by Fock-state support restricted to even or odd photon numbers~\cite{DodonovMalkinManko1974}. The present states do not generally possess such strict even/odd photon-number selection. Their cat-like character instead arises from multicomponent phase-space structure, Wigner negativity, and quantum interference organized by the residual \(\mathbb{Z}_{n+1}\) symmetry.

The present work is motivated in part by the \(n=1\) quantum model of degenerate dual-pump spontaneous four-wave mixing in a \(\chi^{(3)}\) microring resonator~\cite{SinghTeretenkov2026}. In that problem, the diagonal Kerr contribution can be separated by an interaction-picture transformation generated by the SPM and XPM Hamiltonian. The resulting representation highlights the resonant photon-conversion process while retaining the nonlinear Kerr phases in transformed form. Here we extend this viewpoint to the \(n=2\) and \(n=3\) intraresonance branches and compare an effective photon-conversion model with the full Kerr Hamiltonian containing SPM, XPM, and FWM terms.

The paper is organized as follows. Section~\ref{sec:hamiltonian} introduces the multimode Kerr Hamiltonian and its explicit four-mode \(n=2\) decomposition. Section~\ref{sec:reduced_model} explains the interaction-picture separation of diagonal Kerr phases and defines the effective reduced photon-conversion Hamiltonian. Section~\ref{sec:sym_phase} derives the \(\mathbb{Z}_{n+1}\) symmetry and the \(n=2\) phase-locking equations. Section~\ref{sec:models} lists the reduced Hamiltonians for \(n=1\), \(n=2\), and \(n=3\). Section~\ref{sec:dynamics} defines the quantum dynamics and the normalized interaction length \(\tau=g_2t\). Section~\ref{sec:diagnostics} introduces the diagnostics used at \(\tau_{\max}\). Section~\ref{sec:results} presents the quantum-depleted numerical results for the representative modes \(a_0\) and \(a_1\) and compares the reduced and full Kerr models.

\section{Multimode Kerr Hamiltonian}
\label{sec:hamiltonian}

We consider an intraresonance frequency comb generated inside a doubly pumped Kerr microresonator. The optical modes are indexed by
\begin{equation}
    \mu=0,1,\ldots,n+1,
\end{equation}
where the edge modes \(\mu=0\) and \(\mu=n+1\) are externally pumped, while the intermediate modes \(\mu=1,\ldots,n\) are generated through resonant four-wave mixing.

The conservative multimode Kerr Hamiltonian can be written in the standard four-wave-mixing form~\cite{Chembo2016}
\begin{equation}
    \hat H_{\rm Kerr}
    =
    \frac{\hbar g_n}{2}
    \sum_{i+j=k+l}
    \hat a_i^{\dagger}\hat a_j^{\dagger}\hat a_k\hat a_l .
    \label{eq:Hfull}
\end{equation}
The selection rule \(i+j=k+l\) expresses conservation of the frequency index in resonant four-wave mixing. The sum runs over ordered index tuples satisfying the selection rule. Hence, if the ordered constrained sum is rewritten as a sum
over distinct normally ordered monomials, the monomial
\(\hat a_i^{\dagger}\hat a_j^{\dagger}\hat a_k\hat a_l\)
appears with the combinatorial multiplicity $
(2-\delta_{ij})
(2-\delta_{kl})$. The symbol \(g_n\) denotes the nonlinear coupling scale, e.g. \(g_2\) is used for the \(n=2\) branch and for defining the normalized interaction length below. 

\subsection{Index-phase symmetry of the Kerr Hamiltonian}

Consider the transformation
\begin{equation}
    \hat a_\mu \longrightarrow e^{\ii \beta \mu}\hat a_\mu .
    \label{eq:continuous_symmetry}
\end{equation}
Each monomial transforms as
\begin{align}
    \hat a_i^{\dagger}\hat a_j^{\dagger}\hat a_k\hat a_l
    &\longrightarrow
    e^{-\ii\beta i}e^{-\ii\beta j}e^{\ii\beta k}e^{\ii\beta l}
    \hat a_i^{\dagger}\hat a_j^{\dagger}\hat a_k\hat a_l
    \\
    &=
    e^{\ii\beta(-i-j+k+l)}
    \hat a_i^{\dagger}\hat a_j^{\dagger}\hat a_k\hat a_l .
\end{align}
Because \(i+j=k+l\), the phase factor is unity. Therefore \(\hat H_{\rm Kerr}\) is invariant under the continuous index-phase transformation~\eqref{eq:continuous_symmetry}. For the quantum initial-value problem considered below, the chosen coherent
pump phases select a residual discrete subgroup of this continuous symmetry.

\subsection{Explicit \(n=2\) decomposition into SPM, XPM, and FWM terms}
\label{sec:n2_full_kerr}

For the \(n=2\) intraresonance branch, the four modes are \(\hat a_0,\hat a_1,\hat a_2,\hat a_3\), where \(\hat a_0\) and \(\hat a_3\) are the pump modes and \(\hat a_1,\hat a_2\) are generated modes. The full Kerr Hamiltonian can be decomposed as
\begin{equation}
    \hat H_{\rm Kerr}^{(n=2)}
    =
    \hat H_{\rm SPM}
    +
    \hat H_{\rm XPM}
    +
    \hat H_{\rm FWM} .
    \label{eq:Hkerr_decomp_n2}
\end{equation}
The self-phase-modulation part is
\begin{equation}
    \hat H_{\rm SPM}
    =
    \frac{\hbar g_2}{2}
    \sum_{j=0}^{3}
    \hat a_j^{\dagger2}\hat a_j^2 .
    \label{eq:Hspm_n2}
\end{equation}
The cross-phase-modulation part is
\begin{equation}
    \hat H_{\rm XPM}
    =
    2\hbar g_2
    \sum_{0\leq i<j\leq3}
    \hat n_i\hat n_j,
    \qquad
    \hat n_j=\hat a_j^\dagger\hat a_j .
    \label{eq:Hxpm_n2}
\end{equation}
The photon-conversion part is
\begin{align}
    \hat H_{\rm FWM}
    =
    \hbar g_2\Big[&
    \hat a_1^{\dagger2}\hat a_0\hat a_2
    +
    \hat a_2^{\dagger2}\hat a_1\hat a_3
    +
    2\hat a_1^\dagger\hat a_2^\dagger\hat a_0\hat a_3
    +
    {\rm H.c.}
    \Big].
    \label{eq:Hfwm_n2_compact}
\end{align}
The three conversion channels are
\begin{align}
    \omega_0+\omega_2&=2\omega_1,\label{eq:ch1}\\
    \omega_1+\omega_3&=2\omega_2,\label{eq:ch2}\\
    \omega_0+\omega_3&=\omega_1+\omega_2.\label{eq:ch3}
\end{align}
The factor of \(2\) in the nondegenerate process \(\hat a_1^\dagger\hat a_2^\dagger\hat a_0\hat a_3\) arises from equivalent permutations in the constrained Kerr sum. The diagonal terms \(\hat H_{\rm SPM}\) and \(\hat H_{\rm XPM}\) do not create photons in new modes. They generate intensity-dependent nonlinear frequency shifts. The off-diagonal terms \(\hat H_{\rm FWM}\) are responsible for photon conversion, pump depletion, and the phase-locking network.

\section{Interaction-picture separation of diagonal Kerr phases}
\label{sec:reduced_model}

The full Kerr Hamiltonian can be decomposed into diagonal and photon-conversion contributions,
\begin{equation}
\hat H_{\rm Kerr}
=
\hat H_{\rm diag}
+
\hat H_{\rm FWM},
\label{eq:Hkerr_split}
\end{equation}
where
\begin{equation}
\hat H_{\rm diag}
=
\hat H_{\rm SPM}
+
\hat H_{\rm XPM}
\label{eq:Hdiag}
\end{equation}
contains only photon-number operators. This structure naturally motivates the exact interaction-picture transformation
\begin{equation}
\hat U_0(t)
=
\exp
\!\left[
-\frac{\ii}{\hbar}
\hat H_{\rm diag} t
\right],
\label{eq:U0}
\end{equation}
under which the interaction Hamiltonian becomes
\begin{equation}
\hat H_I(t)
=
\hat U_0^\dagger(t)
\hat H_{\rm FWM}
\hat U_0(t).
\label{eq:HI}
\end{equation}
The transformation removes the diagonal Kerr contribution from the explicit Schr\"odinger equation but does not eliminate its physical influence. Instead, SPM and XPM reappear as operator-valued phase factors dressing the photon-conversion processes. Consequently, the nonlinear Kerr shifts cannot, in general, be interpreted as simple c-number detunings.

For the three-mode dual-pump spontaneous four-wave-mixing model considered in Ref.~\cite{SinghTeretenkov2026}, the interaction-picture representation leads to a particularly simple resonant conversion Hamiltonian. In contrast, the present intraresonance \(n=2\) and \(n=3\) systems contain several simultaneously resonant four-wave-mixing channels. Whether the corresponding operator-valued Kerr phases can be simplified in a manner analogous to the three-mode case remains an open problem.

Motivated by this interaction-picture structure, we introduce the effective resonant photon-conversion Hamiltonian
\begin{equation}
\hat H_{\rm red}
=
\hat H_{\rm FWM},
\label{eq:Hred}
\end{equation}
which isolates the coherent conversion network responsible for quantum pump depletion, entanglement generation, and the emergence of discrete \(\mathbb{Z}_{n+1}\)-symmetric phase-space structures.

For the \(n=2\) branch,
\begin{equation}
\hat H_{\rm red}^{(n=2)}
=
\hbar g_2
\left(
\hat a_1^{\dagger 2}\hat a_0\hat a_2
+
\hat a_2^{\dagger 2}\hat a_1\hat a_3
+
2\hat a_1^\dagger
\hat a_2^\dagger
\hat a_0
\hat a_3
+
{\rm H.c.}
\right).
\label{eq:Hred_n2}
\end{equation}
The corresponding \(n=3\) Hamiltonian contains the seven resonant photon-conversion channels discussed in Sec.~\ref{sec:models}.

Equation~\eqref{eq:Hred_n2} should be viewed as an effective resonant model rather than an exact reduction of the full quantum Kerr dynamics. Its role is to reveal the coherent photon-conversion processes and the underlying \(\mathbb{Z}_{n+1}\) symmetry responsible for the formation of multicomponent Schr\"odinger-cat-like states. The complete Kerr Hamiltonian, with SPM and XPM retained, is then used to quantify how nonlinear phase accumulation modifies, distorts, or suppresses these interference structures.

In a semiclassical description, the diagonal Kerr contribution may be interpreted in terms of effective nonlinear detunings associated with individual conversion channels. Such quantities provide useful physical intuition regarding nonlinear phase mismatch. However, within the full quantum theory they correspond to operator-valued phase shifts and therefore should not be regarded as exact compensation conditions.

\section{Discrete \texorpdfstring{\(\mathbb{Z}_{n+1}\)}{Z(n+1)} symmetry and phase-locking mechanisms}
\label{sec:sym_phase}

The two external pumps fix the phases of modes \(\mu=0\) and
\(\mu=n+1\). In the fully quantum simulations considered here, this
does not mean that the pumps are kept as prescribed classical fields during
the evolution. Rather, the pump modes are included as quantum harmonic
oscillators, and their coherent amplitudes are fixed only at the level of the
initial condition,
\begin{equation}\label{eq:initial_state_general}
    \ket{\Psi(0)}
    =
    \ket{\alpha_0}_{0}
    \ket{0}_{1}\cdots
    \ket{0}_{n}
    \ket{\alpha_{n+1}}_{n+1},
\end{equation}
with prescribed complex phases of \(\alpha_0\) and \(\alpha_{n+1}\). The Kerr Hamiltonian itself remains invariant under the continuous
index-phase transformation~\eqref{eq:continuous_symmetry}. 
However, this transformation maps the initial pump amplitude
\(\alpha_{n+1}\) to \(e^{\ii\beta(n+1)}\alpha_{n+1}\), while
\(\alpha_0\) is unchanged. Thus, the same quantum initial-value problem is
invariant only if
\begin{equation}
    e^{\ii\beta(n+1)}=1 ,
\end{equation}
i.e.
\begin{equation}
    \beta_s=\frac{2\pi s}{n+1},
    \qquad
    s=0,1,\ldots,n .
\end{equation}
The continuous index-phase symmetry of the Hamiltonian is therefore reduced,
by the chosen coherent pump phases in the quantum initial condition, to the
residual cyclic group
\begin{equation}
    \mathbb{Z}_{n+1}:
    \qquad
    \hat a_\mu
    \longrightarrow
    e^{\ii 2\pi s\mu/(n+1)}\hat a_\mu .
    \label{eq:Zn_symmetry}
\end{equation}
This group determines the number of symmetry-related phase branches and
therefore suggests the number of dominant phase-space components that may
appear in cat-like Wigner structures when coherent interference between the
branches is preserved. In the reduced photon-conversion model, we show below
that the quantum-depleted dynamics can preserve enough coherence between these
branches to produce Wigner interference fringes and negativity.

The same $\mathbb{Z}_{n+1}$ structure can be seen at the level of classical
phase locking. Writing the classical field amplitudes as
\begin{equation}
a_\mu=A_\mu e^{\ii\phi_\mu},
\end{equation}
a resonant FWM term
$\hat a_i^\dagger \hat a_j^\dagger \hat a_k\hat a_l$ is phase matched when
\begin{equation}
\phi_i+\phi_j-\phi_k-\phi_l=0
\quad (\mathrm{mod}\;2\pi),
\qquad i+j=k+l .
\end{equation}
Thus the phase increments between neighboring frequency-index modes must be
consistent with the additive index-selection rule. A phase-locked branch is
therefore described by an affine dependence of the phase on the mode index,
\begin{equation}
\phi_\mu=\phi_0+\mu\Delta\phi .
\end{equation}
For fixed pump phases, the second pump at $\mu=n+1$ imposes
\begin{equation}
\phi_{n+1}-\phi_0=(n+1)\Delta\phi=2\pi s
\quad (\mathrm{mod}\;2\pi),
\end{equation}
or
\begin{equation}
\Delta\phi_s=\frac{2\pi s}{n+1},
\qquad s=0,1,\ldots,n .
\label{eq:phase_spacing}
\end{equation}
These \(n+1\) possible phase increments are precisely the classical
phase-locking counterparts of the residual $\mathbb{Z}_{n+1}$ transformations.

For \(n=2\), the three conversion phases are
\begin{align}
    \Phi_1&=\theta_0+\theta_3-\theta_1-\theta_2,\\
    \Phi_2&=\theta_0+\theta_2-2\theta_1,\\
    \Phi_3&=\theta_1+\theta_3-2\theta_2 .
\end{align}
Constructive phase locking gives
\begin{align}
    \theta_1&=
    \frac{2\theta_0+\theta_3}{3}
    +\frac{2\pi q}{3},
    \label{eq:theta1_solution}\\
    \theta_2&=
    \frac{\theta_0+2\theta_3}{3}
    -\frac{2\pi q}{3},
    \label{eq:theta2_solution}
\end{align}
with \(q=0,1,2\). Thus the pump phases set the central direction, while \(q\) labels the three \(\mathbb{Z}_3\)-related branches. This is the group-theoretic origin of three-component cat-like structures in the reduced \(n=2\) quantum model.

\section{Reduced quantum models for \texorpdfstring{\(n=1\), \(n=2\), and \(n=3\)}{n=1, n=2, and n=3}}
\label{sec:models}

For \(n=1\), the modes are \(a_0,a_1,a_2\), with \(a_0\) and \(a_2\) pumped and \(a_1\) generated. The dominant conversion Hamiltonian is
\begin{equation}
    \hat H_{n=1}
    =
    \hbar g_1
    \left(
    \hat a_1^{\dagger 2}\hat a_0\hat a_2
    +
    \hat a_0^{\dagger}\hat a_2^{\dagger}\hat a_1^2
    \right),
    \label{eq:Hn1}
\end{equation}
with residual \(\mathbb{Z}_2\) symmetry.

For \(n=2\), the reduced Hamiltonian is
\begin{align}
    \hat H_{n=2}
    =\hbar g_2\Big(&
    \hat a_1^{\dagger 2}\hat a_0\hat a_2
    +
    \hat a_2^{\dagger 2}\hat a_1\hat a_3
    +2\hat a_1^{\dagger}\hat a_2^{\dagger}\hat a_0\hat a_3
    +\mathrm{H.c.}
    \Big),
    \label{eq:Hn2}
\end{align}
with residual \(\mathbb{Z}_3\) symmetry.

For \(n=3\), the modes are \(a_0,a_1,a_2,a_3,a_4\), with pumps at \(a_0\) and \(a_4\). The photon-conversion Hamiltonian is
\begin{equation}
\begin{aligned}
\hat H_{\rm FWM}^{(n=3)}
=\hbar g_3\Big[&
\hat a_1^{\dagger2}\hat a_0\hat a_2
+2\hat a_1^\dagger\hat a_2^\dagger\hat a_0\hat a_3
+2\hat a_1^\dagger\hat a_3^\dagger\hat a_0\hat a_4
\\
&+
\hat a_2^{\dagger2}\hat a_0\hat a_4
+\hat a_2^{\dagger2}\hat a_1\hat a_3
+2\hat a_2^\dagger\hat a_3^\dagger\hat a_1\hat a_4
\\
&+
\hat a_3^{\dagger2}\hat a_2\hat a_4
+\mathrm{H.c.}\Big].
\end{aligned}
\label{eq:Hn3}
\end{equation}
With fixed coherent pump phases, the corresponding initial-value problem has
a residual \(\mathbb{Z}_4\) symmetry, suggesting  four dominant phase-space components in the ideal compensated quantum-depleted model.

\section{Quantum dynamics}
\label{sec:dynamics}

We use the dimensionless interaction length
\begin{equation}
    \tau = g_2 t ,
    \label{eq:tau_definition}
\end{equation}
where \(g_2\) is the characteristic four-wave-mixing coupling strength of the \(n=2\) intraresonance branch. In these normalized units the nondissipative evolution is written as
\begin{equation}
    \frac{\dd \rho}{\dd \tau}
    =
    -\frac{\ii}{\hbar g_2}
    [\hat H,\rho].
    \label{eq:vonNeumann_tau}
\end{equation}
For the reduced Hamiltonian \(\hat H=\hat H_{\rm red}^{(n=2)}\), this is equivalent to setting \(g_2=1\) in the numerical simulations. Throughout this work, we set $\hbar = 1$.

Both the reduced photon-conversion Hamiltonian and the full Kerr Hamiltonian may be studied in dissipative settings. Here we focus on unitary dynamics because it most directly reveals the coherent interference effects associated with quantum pump depletion, Wigner-function negativity, and the discrete \(\mathbb{Z}_{n+1}\) phase-space structure. Dissipation and uncompensated Kerr phase diffusion both tend to suppress interference fringes, and their combined influence is left for future work.

For the representative \(n=2\) simulations, the initial state \eqref{eq:initial_state_general} takes the form
\begin{equation}
    \ket{\Psi(0)}
    =
    \ket{\alpha_0}_{0}
    \ket{0}_{1}
    \ket{0}_{2}
    \ket{\alpha_3}_{3}.
    \label{eq:initial_state}
\end{equation}
In numerical computations we assume
\begin{equation}
    |\alpha_0|^2
    =
    |\alpha_3|^2
    =
    9 .
\end{equation}
The pump modes \(a_0\) and \(a_3\) are initialized in coherent states, while the generated modes \(a_1\) and \(a_2\) are initialized in vacuum states. The pump modes are therefore treated as quantum modes (for comparison, a brief analysis of the approximate model with an undepleted
pump can be found in Appendix~\ref{app:undepleted}). Pump depletion, back-action, entanglement generation, and non-Gaussian reduced states emerge dynamically rather than being imposed through classical undepleted pump amplitudes.

\section{Diagnostics at \texorpdfstring{\(\tau_{\max}\)}{tau max}}
\label{sec:diagnostics}

The characteristic interaction length is chosen as
\begin{equation}
    \tau_{\max}
    =
    g_2 t_{\max},
    \qquad
    t_{\max}
    =
    \arg\max_{t}\langle \hat n_1(t)\rangle .
    \label{eq:taumax}
\end{equation}
Equivalently, in the normalized simulations,
\begin{equation}
    \tau_{\max}
    =
    \arg\max_{\tau}
    \langle \hat n_1(\tau)\rangle .
\end{equation}
For the parameter range considered here, this point also lies close to the minimum of the pump population \(\langle n_0\rangle\) and to the strongest Wigner negativity of the representative generated mode. Thus \(\tau_{\max}\) marks the stage at which the pump-to-generated-mode conversion and the cat-like phase-space structure are most pronounced.

For each mode \(\mu\), the reduced density operator is obtained by tracing over all other modes,
\begin{equation}
    \rho_\mu(\tau_{\max})
    =
    \Tr_{\{\nu:\nu\neq\mu\}}
    \left[
    \rho(\tau_{\max})
    \right].
    \label{eq:reducedrho}
\end{equation}
The notation \(\Tr_{\{\nu:\nu\neq\mu\}}\) means that all modes except \(\mu\) are traced out. For example, in the four-mode \(n=2\) system, \(\rho_1=\Tr_{0,2,3}\rho\).

The photon-number distribution of mode \(\mu\) is
\begin{equation}
P_\mu(n)
=
\langle n|
\rho_\mu
|n\rangle .
\label{eq:photon_distribution}
\end{equation}
These distributions are used to distinguish the present noncanonical cat-like states from the canonical even and odd coherent states. In particular, the states generated here are not restricted to purely even or purely odd photon-number sectors.

The single-mode Wigner function of mode \(\mu\) is defined in the quadrature representation as
\begin{equation}
W_\mu(x,p)
=
\frac{1}{2\pi}
\int_{-\infty}^{\infty}
\dd y\,
e^{-\ii p y}
\left\langle
x+\frac{y}{2}
\right|
\rho_\mu
\left|
x-\frac{y}{2}
\right\rangle .
\label{eq:wigner_integral}
\end{equation}
Here \(x\) and \(p\) are the dimensionless quadratures,
\begin{equation}
\hat X_\mu
=
\frac{\hat a_\mu+\hat a_\mu^\dagger}{\sqrt2},
\qquad
\hat P_\mu
=
\frac{\hat a_\mu-\hat a_\mu^\dagger}{\ii\sqrt2},
\end{equation}
and \(\ket{x}\) denotes an eigenstate of \(\hat X_\mu\). Equivalently, the same Wigner function can be written in the displaced-parity form
\begin{equation}
W_\mu(x,p)
=
\frac{1}{\pi}
\Tr
\left[
\rho_\mu
\hat D(\alpha)
\hat \Pi
\hat D^\dagger(\alpha)
\right],
\label{eq:wigner_parity}
\end{equation}
where
\begin{equation}
\hat D(\alpha)
=
\exp
\left(
\alpha \hat a_\mu^\dagger
-
\alpha^* \hat a_\mu
\right)
\end{equation}
is the displacement operator,
\begin{equation}
\hat \Pi
=
\exp
\left(
\ii\pi
\hat a_\mu^\dagger\hat a_\mu
\right)
\end{equation}
is the parity operator, and
\begin{equation}
\alpha=\frac{x+\ii p}{\sqrt2}.
\end{equation}
The quadrature marginals are
\begin{align}
    P_\mu(x)
    &=
    \int_{-\infty}^{\infty}
    W_\mu(x,p)\,\dd p,
    \\
    P_\mu(p)
    &=
    \int_{-\infty}^{\infty}
    W_\mu(x,p)\,\dd x .
\end{align}
The minimum value
\begin{equation}
    W_{\min}^{(\mu)}(\tau)
    =
    \min_{x,p} W_\mu(x,p;\tau)
\end{equation}
is used as a compact indicator of Wigner negativity. Negative values of \(W_{\min}^{(\mu)}\) indicate nonclassical interference and are central to identifying Schr\"odinger-cat-like behavior.

Photon-number fluctuations are quantified by the Fano factor
\begin{equation}
    F_\mu
    =
    \frac{
    \langle \hat n_\mu^2\rangle
    -
    \langle \hat n_\mu\rangle^2
    }{
    \langle \hat n_\mu\rangle
    }.
    \label{eq:fano}
\end{equation}
Values \(F_\mu>1\) indicate super-Poissonian statistics, while \(F_\mu<1\) indicates sub-Poissonian statistics. For a coherent state,
\begin{equation}
    \Delta X_\mu^2
    =
    \Delta P_\mu^2
    =
    \frac12 .
\end{equation}
Quadrature squeezing is therefore present whenever either variance falls below \(1/2\).

Finally, for a pure global state, the entanglement between mode \(\mu\) and the remaining modes is quantified through the single-mode Schmidt number
\begin{equation}
    K_{\mu|\bar\mu}
    =
    \frac{1}{\Tr(\rho_\mu^2)} .
    \label{eq:schmidt}
\end{equation}
Here \(K_{\mu|\bar\mu}=1\) corresponds to a separable mode, while \(K_{\mu|\bar\mu}>1\) indicates entanglement between mode \(\mu\) and the rest of the system.

\section{Quantum-depleted dynamics and non-Gaussian state generation}
\label{sec:results}

We now present numerical simulations of the four-mode \(n=2\) model computed using QuTiP~\cite{QuTiP1,QuTiP2}. The pump modes \(a_0\) and \(a_3\) are initialized in coherent states, whereas the generated modes \(a_1\) and \(a_2\) start in vacuum. Because of the symmetry of the two pump modes and the correlated generation of the two intermediate modes, the phase-space and number-state patterns of \(a_3\) and \(a_2\) are qualitatively similar to those of \(a_0\) and \(a_1\), respectively. We therefore focus the figures on the representative pump mode \(a_0\) and generated mode \(a_1\). This avoids redundant panels while retaining the essential physics of pump depletion and generated-mode non-Gaussianity.

We first analyze the reduced Hamiltonian \(\hat H_{\rm red}^{(n=2)}\) and then compare with the full Kerr Hamiltonian
\begin{equation}
    \hat H_{\rm full}
    =
    \hat H_{\rm SPM}
    +
    \hat H_{\rm XPM}
    +
    \hat H_{\rm FWM}.
\end{equation}
The reduced model isolates ideal resonant photon conversion, while the full model tests the robustness of the cat-like interference pattern against diagonal Kerr phase accumulation.

\subsection{Pump depletion and quadrature dynamics}

\begin{figure}[t]
\centering
\includegraphics[width=\linewidth]{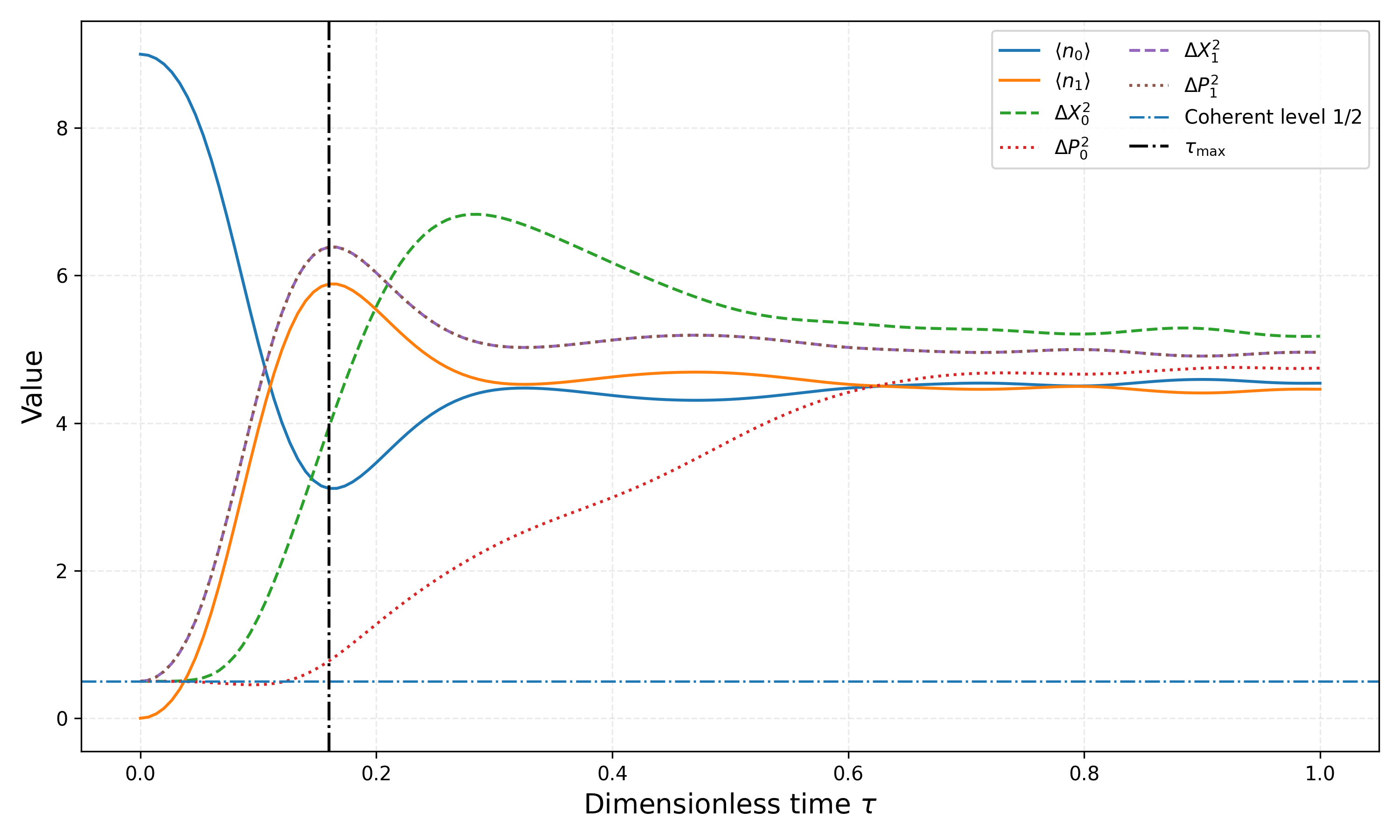}
\caption{
Mean photon numbers \(\langle n_0\rangle\), \(\langle n_1\rangle\) and quadrature variances \(\Delta X_0^2\), \(\Delta P_0^2\), \(\Delta X_1^2\), and \(\Delta P_1^2\) as functions of normalized interaction length \(\tau=g_2t\). The dashed vertical line indicates \(\tau_{\max}\), where the generated-mode population \(\langle n_1\rangle\) is maximal and the pump population \(\langle n_0\rangle\) is near its minimum.
}
\label{fig:mean_variance_a0_a1}
\end{figure}

Figure~\ref{fig:mean_variance_a0_a1} shows the transfer of population from the pump sector to the generated modes. Because the pump modes \(a_0\) and \(a_3\) are treated quantum mechanically, this transfer represents genuine quantum pump depletion rather than an externally prescribed classical drive.

The quadrature variances reveal weak but measurable squeezing of the pump modes. For the representative pump mode \(a_0\), the minimum variance is
\begin{equation}
    \Delta P_0^2 \approx 0.45,
\end{equation}
which occurs at
\begin{equation}
    \tau \simeq 0.09,
\end{equation}
and corresponds to approximately
\begin{equation}
    0.42~{\rm dB}
\end{equation}
of squeezing below the coherent-state level. A similar squeezing behavior is found for the second pump mode \(a_3\), as expected from the symmetry of the two-pump configuration.

No quadrature squeezing is observed for the generated modes \(a_1\) and \(a_2\); their quadrature variances remain at or above the coherent-state level throughout the evolution. Thus the principal nonclassical signature of the generated modes is not quadrature squeezing, but rather the appearance of non-Gaussian Wigner-function interference patterns and negativity near \(\tau=\tau_{\max}\).

\subsection{Photon-number statistics and noncanonical cat-like structure}

\begin{figure}[t]
\centering
\includegraphics[width=\linewidth]{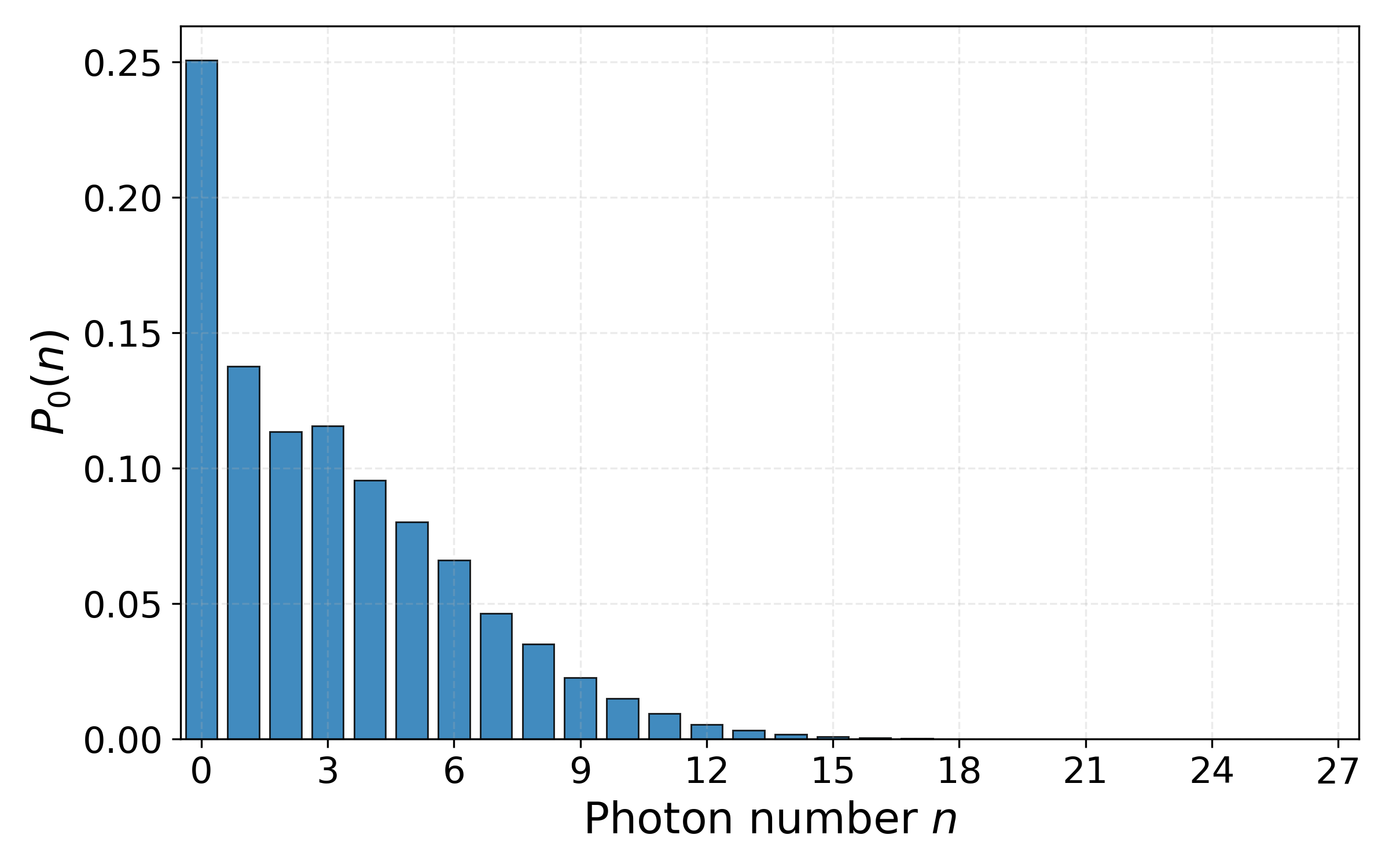}
\caption{
Photon-number distribution \(P_0(n;\tau_{\max})\) of the pump mode \(a_0\). The broadening and deformation relative to the initial coherent-state distribution arise from quantum pump depletion and entanglement between the pump and generated sectors.
}
\label{fig:photon_distribution_a0}
\end{figure}

\begin{figure}[t]
\centering
\includegraphics[width=\linewidth]{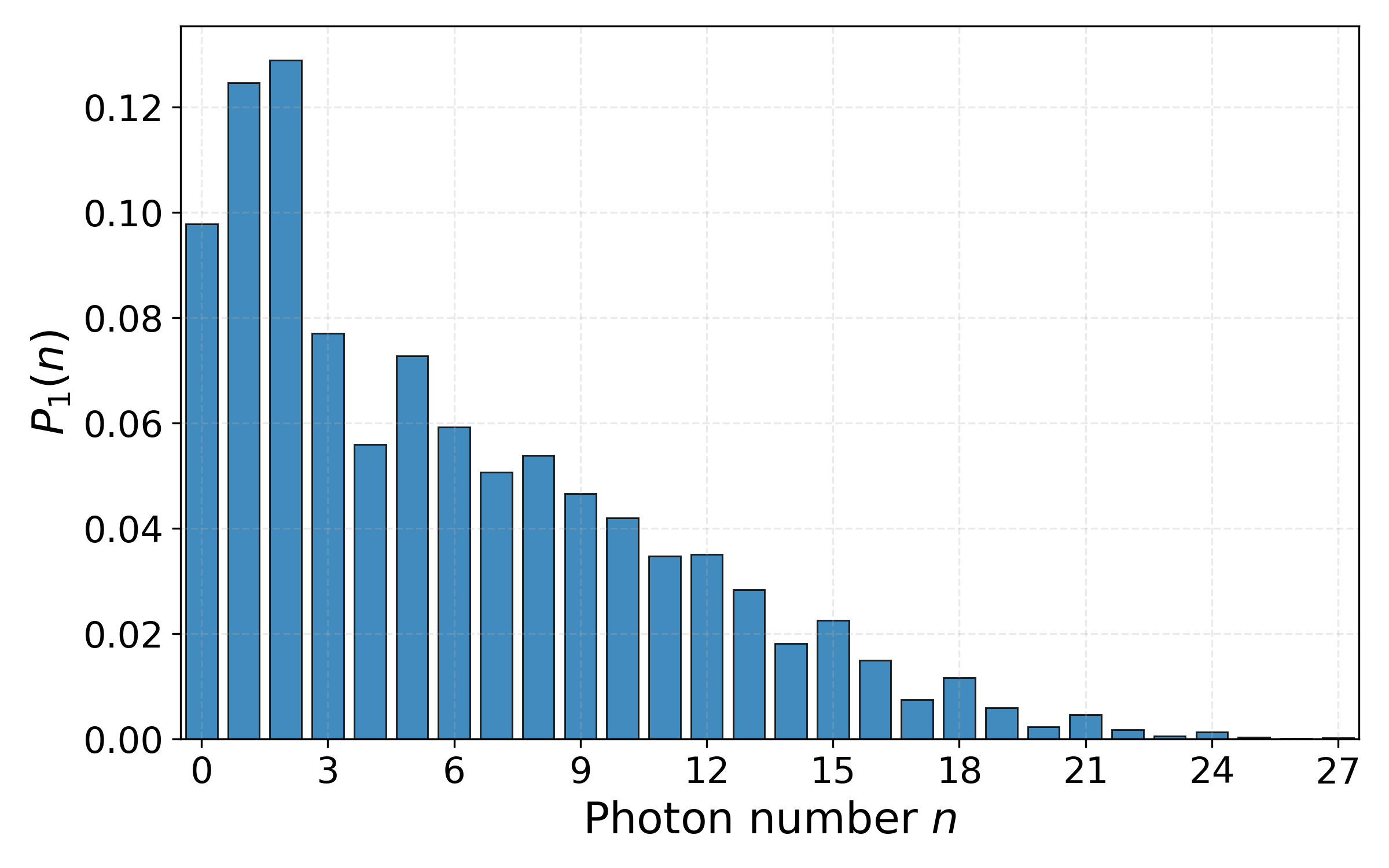}
\caption{
Photon-number distribution \(P_1(n;\tau_{\max})\) of the generated mode \(a_1\). The strongly non-Poissonian statistics demonstrate that the generated field cannot be described as a coherent state and instead reflects the underlying quantum photon-conversion dynamics.
}
\label{fig:photon_distribution_a1}
\end{figure}

Figures~\ref{fig:photon_distribution_a0} and~\ref{fig:photon_distribution_a1} show the photon-number statistics of the representative pump and generated modes at \(\tau=\tau_{\max}\). The distributions deviate substantially from the Poissonian statistics expected for coherent states. Since the photon-number distributions of \(a_3\) and \(a_2\) are qualitatively similar to those of \(a_0\) and \(a_1\), respectively, they are omitted for brevity.

The observed statistics indicate that the generated field is neither coherent nor thermal. It belongs to a strongly correlated multimode quantum state arising from quantum pump depletion and four-wave-mixing interactions. Moreover, the reduced states are not canonical even or odd coherent states: canonical even and odd cat states have support only on even or odd Fock states, respectively~\cite{DodonovMalkinManko1974}. In contrast, the present photon-number distributions generally contain both even and odd photon numbers. The term cat-like is therefore used in a phase-space sense: the states exhibit non-Gaussian multicomponent Wigner structure and interference, but they are not canonical even/odd cats.

\subsection{Wigner signatures of cat-like state formation}

\begin{figure}[t]
\centering
\includegraphics[width=\linewidth]{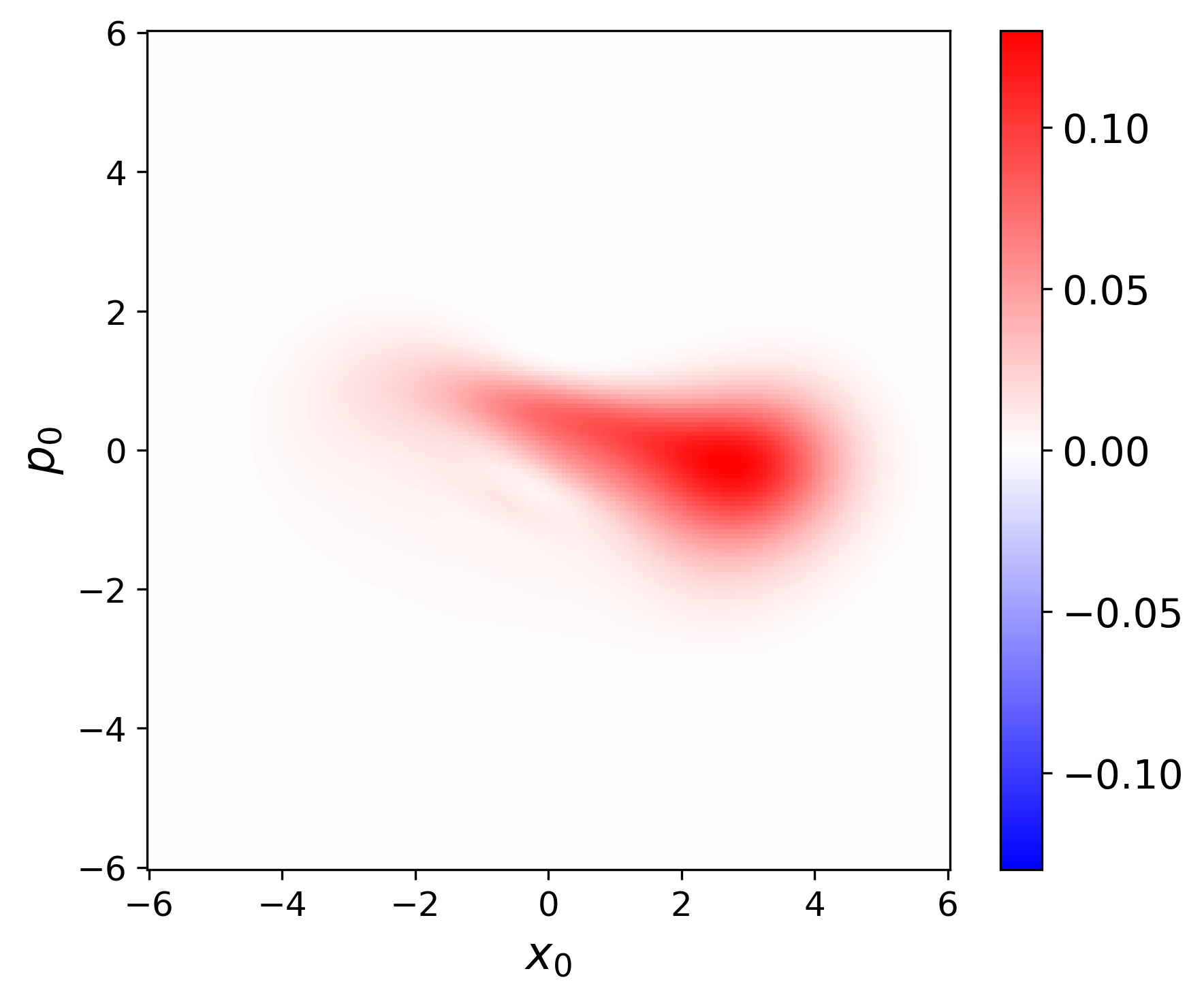}
\caption{
Wigner function \(W_0(x_0,p_0;\tau_{\max})\) of the pump mode \(a_0\) at \(\tau=\tau_{\max}\) for the reduced photon-conversion Hamiltonian. The phase-space distribution departs significantly from the initial coherent-state Gaussian as a consequence of quantum pump depletion and entanglement with the generated modes.
}
\label{fig:wigner_a0}
\end{figure}

\begin{figure}[t]
\centering
\includegraphics[width=\linewidth]{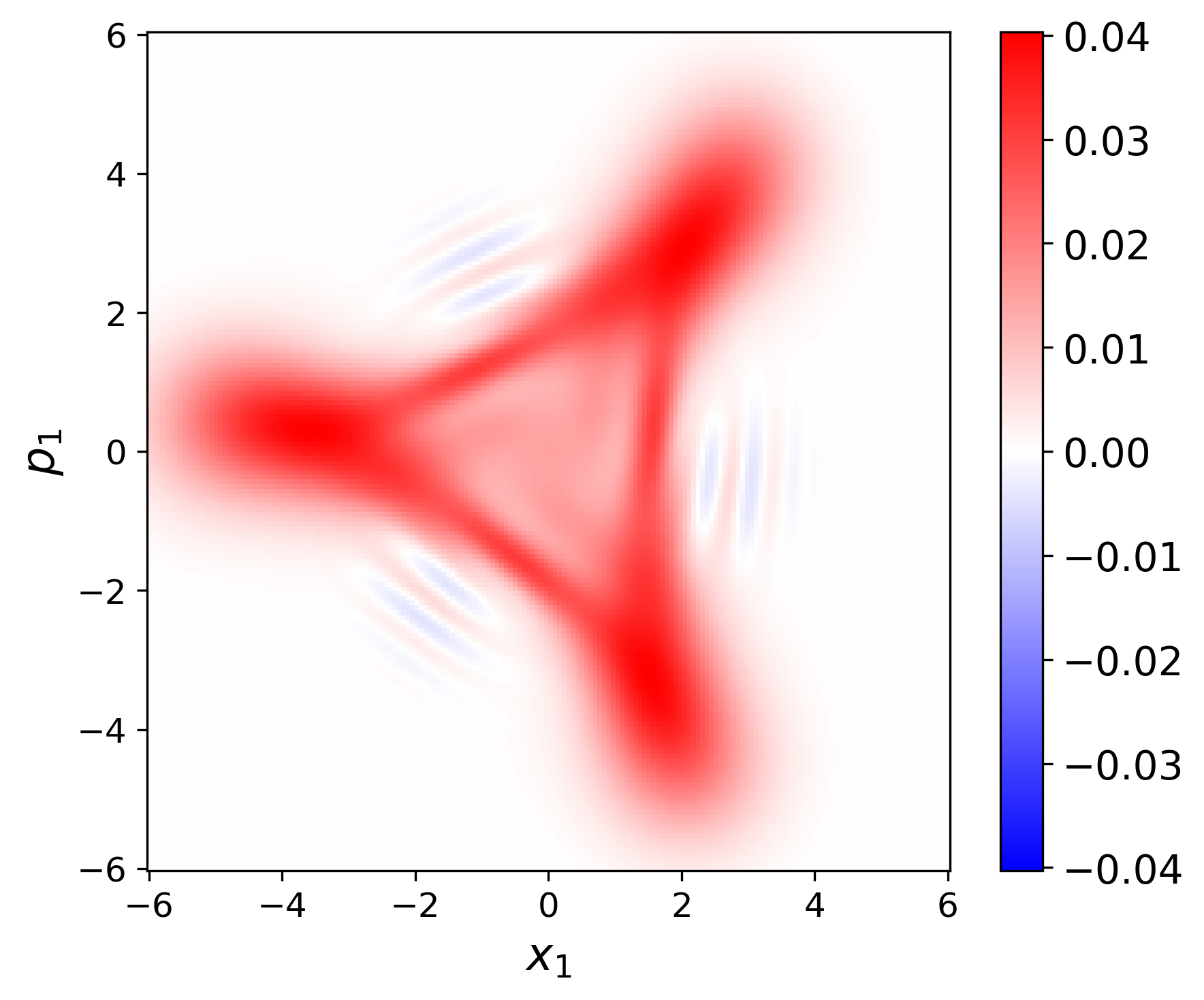}
\caption{
Wigner function \(W_1(x_1,p_1;\tau_{\max})\) of the generated mode \(a_1\) at \(\tau=\tau_{\max}\) for the reduced photon-conversion Hamiltonian. Pronounced interference fringes and negative regions reveal strong nonclassicality and indicate the formation of a Schr\"odinger-cat-like state. The strongest Wigner negativity occurs near \(\tau=\tau_{\max}\), which coincides with the maximum population of \(a_1\) and strong depletion of the pump mode \(a_0\).
}
\label{fig:wigner_a1}
\end{figure}

Figures~\ref{fig:wigner_a0} and~\ref{fig:wigner_a1} show the phase-space structure of the representative pump and generated modes at \(\tau=\tau_{\max}\). The pump mode exhibits strong deviations from the initial coherent-state Gaussian, while the generated mode develops a multimodal interference pattern accompanied by negative regions of the Wigner function. These interference fringes are the features that make cat-like states useful for sensing: small phase-space perturbations can strongly modify the fringe pattern~\cite{SinghTeretenkovPhysicsOpen2024}.

\subsection{Quadrature marginal distributions}

\begin{figure}[t]
\centering
\includegraphics[width=\linewidth]{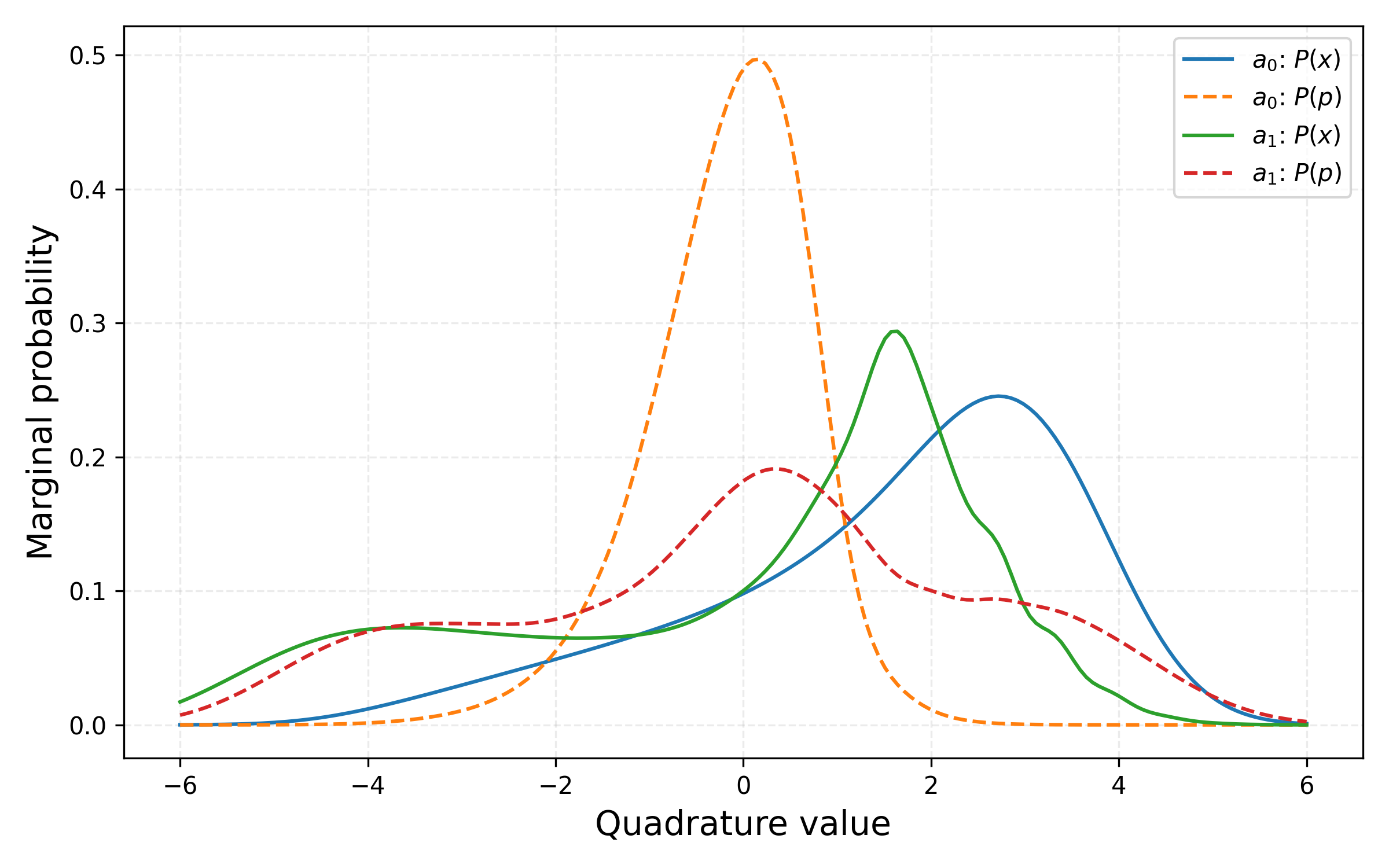}
\caption{
Quadrature marginals \(P(x)\) and \(P(p)\) of the representative modes \(a_0\) and \(a_1\) at \(\tau=\tau_{\max}\).
}
\label{fig:marginals_a0_a1}
\end{figure}

The quadrature marginals in Fig.~\ref{fig:marginals_a0_a1} provide one-dimensional projections of the Wigner functions. Broadening and oscillatory features in these marginals reflect the non-Gaussian structure generated by the photon-conversion dynamics.

\subsection{Photon-number fluctuations and entanglement}

Although separate plots of the Fano factors and Schmidt numbers are not shown, their numerical values are useful for quantifying the strength of the quantum correlations at \(\tau=\tau_{\max}\). For the reduced model, the representative pump and generated modes have
\begin{equation}
    F_0 \approx 2.88,
    \qquad
    F_1 \approx 4.39.
\end{equation}
Both values are well above unity, showing that the photon-number statistics are strongly super-Poissonian and therefore far from coherent-state statistics.

The corresponding single-mode Schmidt numbers are
\begin{equation}
    K_{0|\bar 0} \approx 2.62,
    \qquad
    K_{1|\bar 1} \approx 7.14.
\end{equation}
These values demonstrate significant entanglement between each representative mode and the remaining three modes. The larger value of \(K_{1|\bar 1}\) shows that the generated mode is strongly correlated with the depleted pump sector and with the other generated mode. Thus the observed cat-like Wigner structures are intrinsically multimode states generated by quantum pump depletion and back-action.

\subsection{Effect of the full Kerr Hamiltonian}

\begin{figure}[t]
\centering
\includegraphics[width=\linewidth]{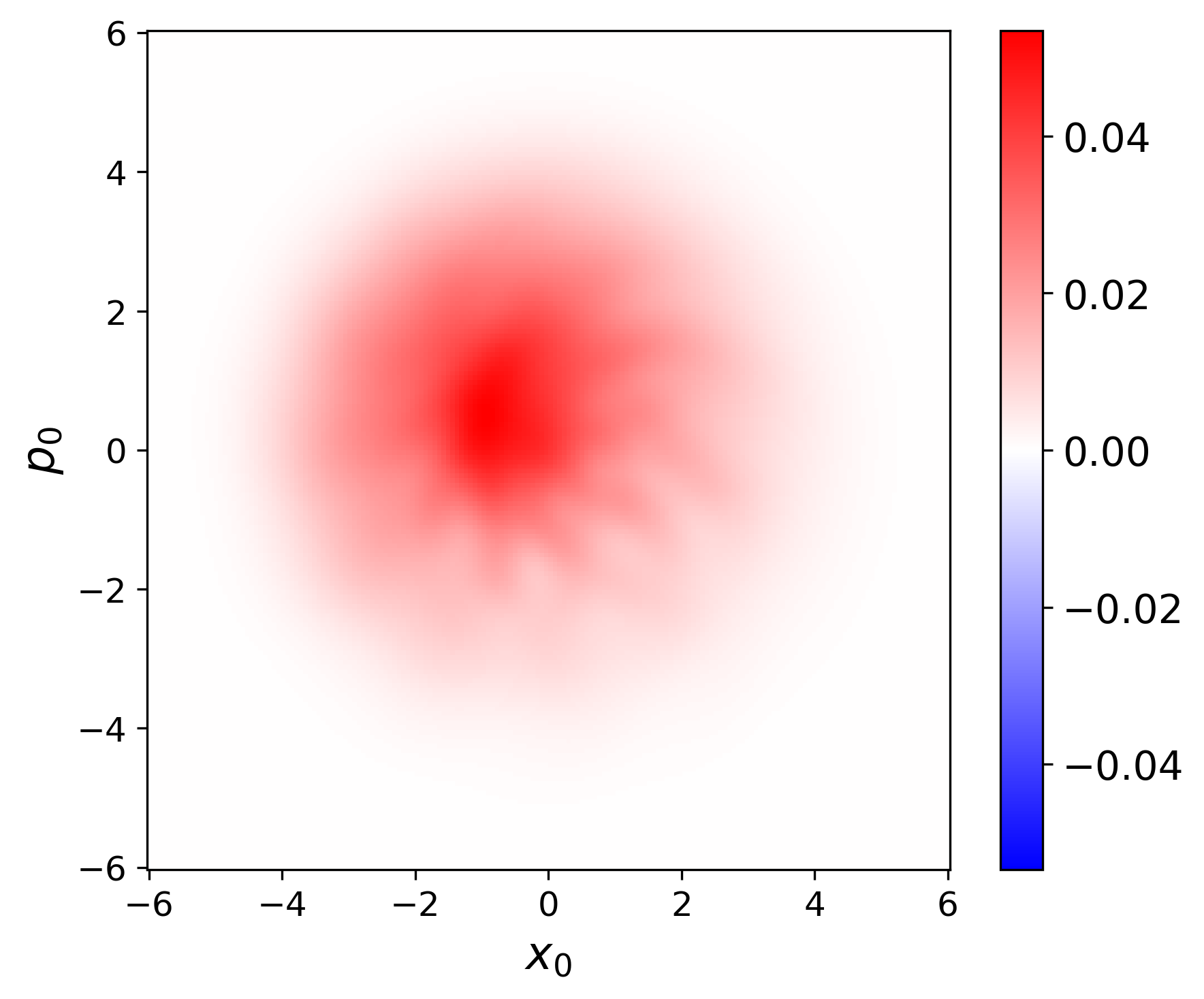}
\caption{
Wigner function \(W_0(x_0,p_0;\tau_{\max})\) of the pump mode \(a_0\) obtained from the full Kerr Hamiltonian including SPM and XPM.
}
\label{fig:full_wigner_a0}
\end{figure}

\begin{figure}[t]
\centering
\includegraphics[width=\linewidth]{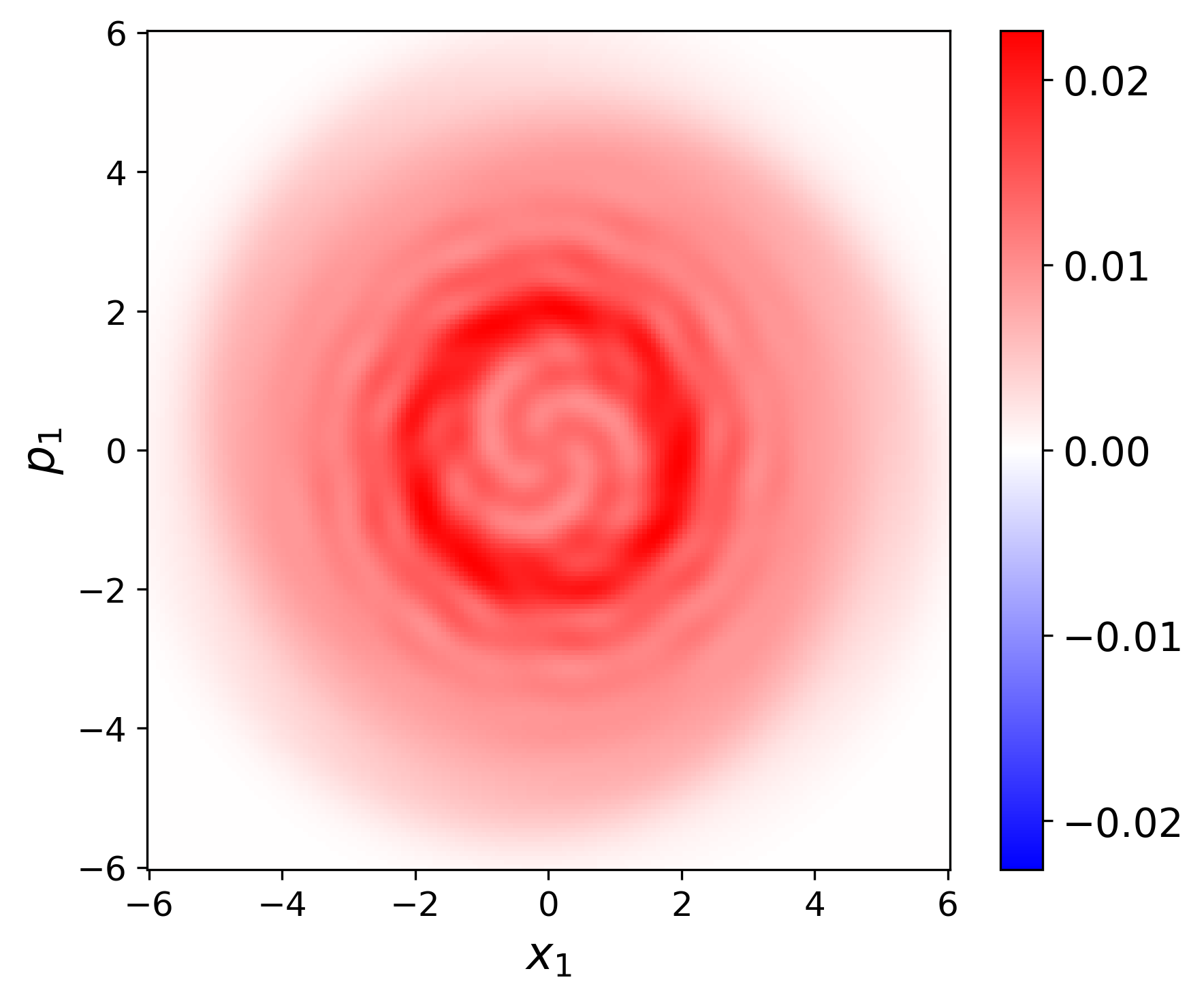}
\caption{
Wigner function \(W_1(x_1,p_1;\tau_{\max})\) of the generated mode \(a_1\) obtained from the full Kerr Hamiltonian. Compared with the reduced model, the interference pattern is strongly weakened and the Wigner function becomes non-negative within numerical resolution, indicating that uncompensated SPM/XPM phase shearing suppresses the clearest cat-like signature.
}
\label{fig:full_wigner_a1}
\end{figure}

When the full Kerr Hamiltonian is used, SPM and XPM generate operator-valued nonlinear phase accumulation. In phase space this appears as shearing and distortion of the Wigner function, as shown in Figs.~\ref{fig:full_wigner_a0} and~\ref{fig:full_wigner_a1}. The resulting states remain non-Gaussian, and the basic quantum statistical quantities, such as photon-number broadening, Fano factors, and single-mode Schmidt numbers, do not change as dramatically as the Wigner negativity. The most sensitive signature is the interference fringe structure itself, which is strongly reduced or eliminated by the full diagonal Kerr dynamics.

\section{Conclusion}

We have developed a quantum-depleted model of intraresonance frequency-comb dynamics in doubly pumped Kerr microresonators. The pump modes are treated quantum mechanically, so pump depletion, back-action, entanglement generation, quadrature fluctuations, and Wigner negativity arise from the same unitary dynamics. The normalized interaction length \(\tau=g_2t\) provides a natural scale for comparing the onset of pump depletion and non-Gaussian state formation.

The central theoretical result is that the four-wave-mixing index-selection
rule, together with the coherent pump phases fixed in the quantum initial
condition, selects a residual cyclic group  \(\mathbb{Z}_{n+1}\). For the \(n=2\) branch this gives \(\mathbb{Z}_3\), and for the \(n=3\) branch it gives \(\mathbb{Z}_4\). The order of this group determines the number of symmetry-related phase
branches and suggests the number of dominant phase-space components that can
appear in the ideal resonant photon-conversion model when coherence between
the branches is preserved.

The cat-like states obtained here are not canonical even or odd coherent states in the sense of Dodonov, Malkin, and Man'ko \cite{DodonovMalkinManko1974}. Their photon-number distributions are not restricted to even or odd Fock sectors. Instead, they are noncanonical multicomponent Schr\"odinger-cat-like states: their defining signatures are multicomponent Wigner structure, Wigner negativity, non-Poissonian statistics, quadrature fluctuations, and entanglement generated by quantum pump depletion.

For the \(n=2\) simulations, the strongest cat-like features occur near \(\tau_{\max}=g_2t_{\max}\), where the generated population \(\langle n_1\rangle\) reaches its maximum and the pump population \(\langle n_0\rangle\) is near its minimum. In the reduced photon-conversion model this point also corresponds to the strongest negative regions of the representative Wigner functions.

At \(\tau_{\max}\), approximately \(65\%\) of the initial pump population has been converted into the generated modes in the reduced FWM model, with
\(
\langle n_0\rangle
=
\langle n_3\rangle
\approx
3.11
\)
and
\(
\langle n_1\rangle
=
\langle n_2\rangle
\approx
5.89.
\)

The quadrature dynamics show modest squeezing in the pump modes \(a_0\) and \(a_3\), rather than in the generated modes. For the representative pump mode \(a_0\), the minimum variance is \(\Delta P_0^2 \approx 0.45\), corresponding to approximately \(0.42\) dB of squeezing below the coherent-state level. No quadrature squeezing is observed for the generated modes \(a_1\) and \(a_2\). Their main nonclassical signature is instead the formation of multicomponent Schr\"odinger-cat-like phase-space structures accompanied by Wigner-function negativity.

We carefully distinguished the effective reduced Hamiltonian from the full Kerr Hamiltonian. The reduced model isolates the coherent photon-conversion network responsible for \(\mathbb{Z}_{n+1}\)-organized cat-like structures. It should not be interpreted as an exact removal of SPM and XPM from the full quantum problem. The diagonal Kerr contribution generates operator-valued nonlinear phases which, in general, cannot be cancelled by ordinary scalar detunings. When the full Kerr Hamiltonian is used, these phases produce shearing of the phase-space distribution. The Wigner functions remain non-Gaussian but become non-negative in the representative simulations, whereas the remaining statistical quantities are modified less dramatically. This shows that Wigner negativity and interference fringes are the most fragile signatures of the cat-like state.

For comparison, Appendix~\ref{app:undepleted} examines the undepleted-pump regime, in which the pump modes are replaced by prescribed classical amplitudes. The generated modes nevertheless develop non-Gaussian Schr\"odinger-cat-like states with Wigner negativity. This occurs because the effective Hamiltonian still contains the cubic photon-conversion terms
$
\alpha_0 \hat a_1^{\dagger 2}\hat a_2
$
and
$
\alpha_3 \hat a_2^{\dagger 2}\hat a_1$
,
which are intrinsically non-Gaussian and can generate Wigner negativity even in the absence of quantum pump depletion. However, the undepleted model cannot describe either pump-generated-mode entanglement or quantum back-action, which remain distinctive features of the fully quantum dynamics investigated in this work.

These results identify quantum-depleted intraresonance Kerr dynamics as a possible route toward generating noncanonical multicomponent non-Gaussian resources for continuous-variable quantum information processing, optical cat-qubit architectures, and quantum sensing \cite{SinghTeretenkovPLA2026}. A key open problem is to determine experimentally accessible regimes in which resonant photon conversion dominates over uncompensated Kerr phase diffusion, allowing the \(\mathbb{Z}_{n+1}\)-organized cat-like structures to survive in realistic microresonators.

\begin{acknowledgments}
R.S. is grateful to Prof. A.V. Masalov for insightful discussions.
\end{acknowledgments}

\clearpage
\appendix
\section{Undepleted-Pump Approximation and Non-Gaussian State Generation}
\label{app:undepleted}
The purpose of this appendix is not to replace the quantum-depleted description developed in the main text, but rather to isolate the role of the nonlinear photon-conversion processes when pump depletion and quantum back-action are neglected.

For comparison with the quantum-depleted model discussed in the main text, it is useful to consider the undepleted-pump approximation. In this limit the pump modes are treated as classical coherent fields and their amplitudes are replaced by c-numbers,
\begin{equation}
    \hat a_0 \rightarrow \alpha_0,
    \qquad
    \hat a_3 \rightarrow \alpha_3 ,
\end{equation}
where \(\alpha_0\) and \(\alpha_3\) are fixed complex amplitudes.

Applying this approximation to the reduced \(n=2\) photon-conversion Hamiltonian yields
\begin{align}
    \hat H_{\rm undepl}
    =
    \hbar g_2 \Big(
    \alpha_0 \hat a_1^{\dagger 2}\hat a_2
    +
    \alpha_3 \hat a_2^{\dagger 2}\hat a_1
    +
    2\alpha_0\alpha_3
    \hat a_1^\dagger \hat a_2^\dagger
    +
    {\rm H.c.}
    \Big).
\end{align}

\begin{figure}
\centering
\includegraphics[width=\linewidth]{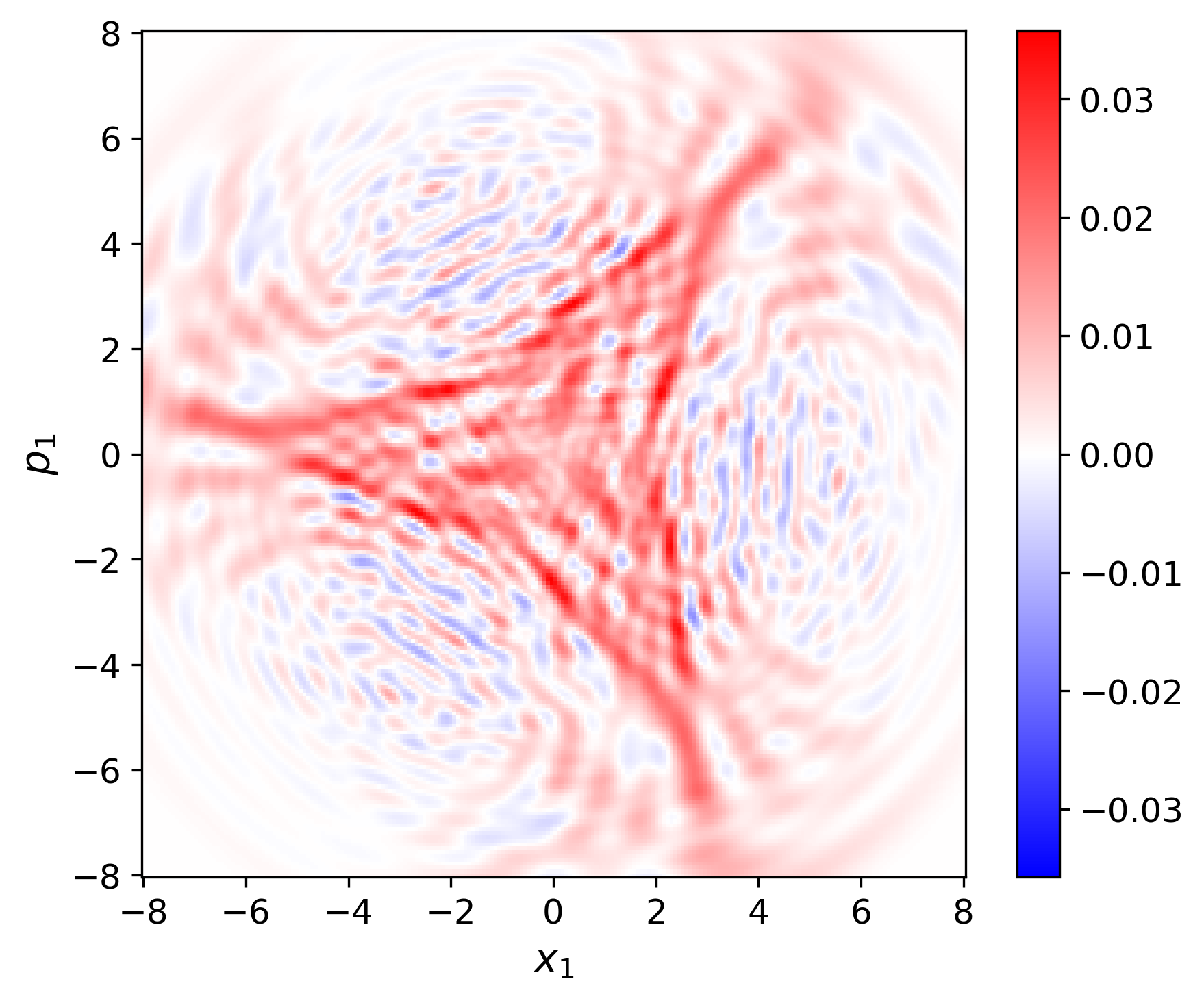}
\caption{Wigner function $W_1(x_1,p_1;\tau_{\max}^{\text{undepl}})$ of the generated mode $a_1$ at $\tau=\tau_{\max}^{\text{undepl}}$, computed under the undepleted-pump approximation for the reduced photon-conversion Hamiltonian---i.e., modes $a_0$ and $a_3$ are treated as externally fixed classical amplitudes. The pronounced interference fringes and negative regions reveal strong nonclassicality and indicate the formation of a Schr\"odinger-cat-like state in mode $a_1$. The strongest Wigner negativity occurs near $\tau_{\max}^{\text{undepl}}$, which coincides with the maximum mean population of the generated mode $a_1$ under these prescribed, nondepleted pump conditions.}
\label{fig:wigneru_a1}
\end{figure}

The generated modes \(a_1\) and \(a_2\) remain quantum degrees of freedom, while the pump amplitudes are externally prescribed and do not evolve dynamically (see Fig.~\ref{fig:wigneru_a1}).
\end{document}